%% file: main.tex
\documentclass[conference]{IEEEtran}
\IEEEoverridecommandlockouts

\AtBeginDocument{%
  \providecommand\BibTeX{{%
    \normalfont B\kern-0.5em{\scshape i\kern-0.25em b}\kern-0.8em\TeX}}}
\usepackage{algorithmic}
\usepackage[color = pink]{todonotes}
 \usepackage[ruled, lined, linesnumbered, commentsnumbered, longend]{algorithm2e}

\SetCommentSty{mycommfont}

\usepackage{xcolor}
\usepackage{ifthen}

\input{colors}
\usepackage{tikz}
\usepackage{pgfplots,pgfplotstable}
\pgfplotsset{
        set layers={
            background,
            main,
        },
    }
\usepackage[numbers]{natbib}
\setcitestyle{square, comma, numbers,sort&compress, super}

\usepackage[T1]{fontenc}
\usepackage{multirow}
\usetikzlibrary{calc, patterns, spy}

\usepackage{listings}
\AtBeginDocument{\DeclareCaptionSubType{lstlisting}}
\usepackage{caption,subcaption}
\usepackage[binary-units=true]{siunitx} %
\usepackage[frozencache,cachedir=minted-cache]{minted}

\usepackage{listings}
\setminted{fontsize=\small,baselinestretch=1}
\usepackage{booktabs}
\usepackage{hyperref}
\usepackage{xspace}

\definecolor{mGreen}{rgb}{0,0.6,0}
\definecolor{mGray}{rgb}{0.5,0.5,0.5}
\definecolor{mPurple}{rgb}{0.58,0,0.82}
\definecolor{backgroundColour}{rgb}{0.95,0.95,0.92}

\lstdefinestyle{CStyle}{
    backgroundcolor=\color{backgroundColour},   
    commentstyle=\color{mGreen},
    keywordstyle=\color{magenta},
    numberstyle=\tiny\color{mGray},
    stringstyle=\color{mPurple},
    basicstyle=\footnotesize,
    breakatwhitespace=false,         
    breaklines=true,                 
    captionpos=b,                    
    keepspaces=true,                 
    numbers=left,                    
    numbersep=5pt,                  
    showspaces=false,                
    showstringspaces=false,
    showtabs=false,                  
    tabsize=2,
    language=C
}

\newcommand{\costmodelframework}{\mbox{CoMoNM}\xspace}

\newcommand{\matvec}{\mbox{\texttt{gemv}}\xspace}
\newcommand{\ourasm}{\mbox{\texttt{llvcnm}}\xspace}

\newcommand{\linalg}{\mbox{\texttt{linalg}}\xspace}

\usepackage{pifont}%

\makeatletter
\lstdefinelanguage{mlir}{
    classoffset=0,
    morekeywords={
        module,
        func,
        cinm,
        cnm,
        cim
    },
    morestring=[b]",
    alsoletter={\%},
    keywordsprefix={\%}
}
\makeatother

\begin{document}

\title{\costmodelframework: A \underline{Co}st \underline{Mo}deling Framework for Compute-\underline{N}ear-\underline{M}emory Systems}

\author{\IEEEauthorblockN{Hamid Farzaneh}
\IEEEauthorblockA{
\textit{TU Dresden, Germany}\\
hamid.farzaneh@tu-dresden.de}
\and
\IEEEauthorblockN{Asif Ali Khan}
\IEEEauthorblockA{
\textit{TU dresden, Germany} \\
asif\_ali.khan@tu-dresden.de}
\and
\IEEEauthorblockN{Jeronimo Castrillon}
\IEEEauthorblockA{
\textit{TU dresden, ScaDS.AI, Germany} \\
jeronimo.castrillon@tu-dresden.de}
}

\maketitle

\begin{abstract}

Compute-Near-Memory (CNM) systems offer a promising approach to mitigate the von Neumann bottleneck by bringing computational units closer to data. However, optimizing for these architectures remains challenging due to their unique hardware and programming models. Existing CNM compilers often rely on manual programmer annotations for offloading and optimizations. Automating these decisions by exploring the optimization space, common in CPU/GPU systems, is difficult for CNMs as constructing and navigating the transformation space %
is tedious and time consuming. 
This is particularly the case during system-level design, where evaluation requires time-consuming simulations.
To address this, we present \emph{\costmodelframework}, a generic cost modeling framework for CNM systems for execution time estimation in milliseconds. 
It takes a high-level, hardware-agnostic application representation, target system specifications, and 
a mapping specification as input and estimates the execution time for the given application on the target CNM system. %
We show how \costmodelframework can be seamlessly integrated into state-of-the-art CNM compilers, %
providing improved offloading decisions.  %
Evaluation on established benchmarks for CNM shows estimation errors within $\pm$7.80\% and $\pm$2.99\%, when compared to the real UPMEM CNM system and Samsung's HBM-PIM simulator. 
Notably, \costmodelframework delivers estimates %
seven orders of magnitude faster compared to the UPMEM and HBM-PIM simulators. %
\end{abstract}

\input{contents/introduction}

\input{contents/new_background}

\input{contents/new_motiv}

\input{contents/costmodelInputs}

\input{contents/costmodel_framwork}
\input{contents/evaluation}

\input{contents/related_work}
\input{contents/conclusion}

\section*{Acknowledgment}
This work is partially funded by the AI competence center ScaDS.AI Dresden/Leipzig (01IS18026A-D), and by the German Research Council (DFG) through the HetCIM project (502388442), and the CRC/TRR 404-Active 3D (528378584).
\sloppy
\bibliographystyle{IEEEtran}
\bibliography{main}

\end{document}

%% file: colors.tex
\definecolor{pairedOneLightBlue}{HTML}{a6cee3}
\definecolor{pairedTwoDarkBlue}{HTML}{1f78b4}
\definecolor{pairedThreeLightGreen}{HTML}{b2df8a}
\definecolor{pairedFourDarkGreen}{HTML}{33a02c}
\definecolor{pairedFiveLightRed}{HTML}{fb9a99}
\definecolor{pairedSixDarkRed}{HTML}{e31a1c}
\definecolor{butter1}{rgb}{0.988,0.914,0.310}
\definecolor{butter2}{rgb}{0.929,0.831,0.000}
\definecolor{butter3}{rgb}{0.769,0.627,0.000}
\definecolor{orange1}{rgb}{0.988,0.686,0.243}
\definecolor{orange2}{rgb}{0.961,0.475,0.000}
\definecolor{orange3}{rgb}{0.808,0.361,0.000}
\definecolor{chocolate1}{rgb}{0.914,0.725,0.431}
\definecolor{chocolate2}{rgb}{0.757,0.490,0.067}
\definecolor{chocolate3}{rgb}{0.561,0.349,0.008}
\definecolor{chameleon1}{rgb}{0.541,0.886,0.204}
\definecolor{chameleon2}{rgb}{0.451,0.824,0.086}
\definecolor{chameleon3}{rgb}{0.306,0.604,0.024}
\definecolor{skyblue1}{rgb}{0.447,0.624,0.812}
\definecolor{skyblue2}{rgb}{0.204,0.396,0.643}
\definecolor{skyblue3}{rgb}{0.125,0.290,0.529}
\definecolor{plum1}{rgb}{0.678,0.498,0.659}
\definecolor{plum2}{rgb}{0.459,0.314,0.482}
\definecolor{plum3}{rgb}{0.361,0.208,0.400}
\definecolor{scarletred1}{rgb}{0.937,0.161,0.161}
\definecolor{scarletred2}{rgb}{0.800,0.000,0.000}
\definecolor{scarletred3}{rgb}{0.643,0.000,0.000}
\definecolor{aluminium1}{rgb}{0.933,0.933,0.925}
\definecolor{aluminium2}{rgb}{0.827,0.843,0.812}
\definecolor{aluminium3}{rgb}{0.729,0.741,0.714}
\definecolor{aluminium4}{rgb}{0.533,0.541,0.522}
\definecolor{aluminium5}{rgb}{0.333,0.341,0.325}
\definecolor{aluminium6}{rgb}{0.180,0.204,0.212}

\definecolor{blind_safe_one_scheme_three_colors}{RGB}{102,194,165}
\definecolor{blind_safe_two_scheme_three_colors}{RGB}{252,141,98}
\definecolor{blind_safe_three_scheme_three_colors}{RGB}{141,160,203}

\definecolor{blind_safe_one_scheme_four_colors}{RGB}{166,206,227}
\definecolor{blind_safe_two_scheme_four_colors}{RGB}{31,120,180}
\definecolor{blind_safe_three_scheme_four_colors}{RGB}{178,223,138}
\definecolor{blind_safe_four_scheme_four_colors}{RGB}{51,160,44}

\definecolor{blind_safe_one_scheme_five_colors}{RGB}{240,249,232}
\definecolor{blind_safe_two_scheme_five_colors}{RGB}{186,228,188}
\definecolor{blind_safe_three_scheme_five_colors}{RGB}{123,204,196}
\definecolor{blind_safe_four_scheme_five_colors}{RGB}{67,162,202}
\definecolor{blind_safe_five_scheme_five_colors}{RGB}{8,104,172} 

\definecolor{blind_safe_one_scheme_seven_colors_grnblu}{RGB}{240,249,232}
\definecolor{blind_safe_two_scheme_seven_colors_grnblu}{RGB}{204,235,197}
\definecolor{blind_safe_three_scheme_seven_colors_grnblu}{RGB}{168,221,181}
\definecolor{blind_safe_four_scheme_seven_colors_grnblu}{RGB}{123,204,196}
\definecolor{blind_safe_five_scheme_seven_colors_grnblu}{RGB}{78,179,211}
\definecolor{blind_safe_six_scheme_seven_colors_grnblu}{RGB}{43,140,190}
\definecolor{blind_safe_seven_scheme_seven_colors_grnblu}{RGB}{8,88,158}

\definecolor{blind_safe_one_scheme_seven_colors}{RGB}{118,42,131}
\definecolor{blind_safe_two_scheme_seven_colors}{RGB}{175,141,195}
\definecolor{blind_safe_three_scheme_seven_colors}{RGB}{231,212,232}
\definecolor{blind_safe_four_scheme_seven_colors}{RGB}{247,247,247}
\definecolor{blind_safe_five_scheme_seven_colors}{RGB}{217,240,211}
\definecolor{blind_safe_six_scheme_seven_colors}{RGB}{127,191,123}
\definecolor{blind_safe_seven_scheme_seven_colors}{RGB}{27,120,55}

\definecolor{yellow_one}{RGB}{255,255,212}
\definecolor{yellow_two}{RGB}{254,217,142}
\definecolor{yellow_three}{RGB}{254,153,41}
\definecolor{yellow_four}{RGB}{217,95,14}
\definecolor{yellow_five}{RGB}{153,52,4}

\definecolor{minted_red_bg}{HTML}{ffdfde}
\definecolor{minted_blue_bg}{HTML}{dcdff7}

\definecolor{minted_ins_norm}{HTML}{ffffff}
\definecolor{minted_ins_mram}{HTML}{facb89}
\definecolor{minted_ins_wram}{HTML}{f5c9e5}
\definecolor{minted_ins_cf}{HTML}{b4e0ab}
\definecolor{minted_ins_comp}{HTML}{afccfa}

\definecolor{codegray}{rgb}{0.5,0.5,0.5}
\definecolor{lightblue}{HTML}{deebff}

\definecolor{one_scheme_seven_colors}{HTML}{c7e9b4}
\definecolor{two_scheme_seven_colors}{HTML}{7fcdbb}
\definecolor{three_scheme_seven_colors}{HTML}{41b6c4}
\definecolor{four_scheme_seven_colors}{HTML}{1d91c0}
\definecolor{five_scheme_seven_colors}{HTML}{225ea8}
\definecolor{six_scheme_seven_colors}{HTML}{253494}
\definecolor{seven_scheme_seven_colors}{HTML}{081d58}

%% file: contents/introduction.tex
\section{Introduction}

The emerging \emph{computing-near-memory} (CNM) paradigm tackles the von-Neumann bottleneck by moving computation closer to the data~\cite{abndp}.
These systems have gained significant attention in recent years due to the growing demand for data-intensive applications domains, such as machine learning and bioinformatics, which expose the limits of conventional computing systems~\cite{hbm, wideio, memristorNew}.
Examples of CNM systems include UPMEM~\cite{UPMEM}, Samsung's HBM-PIM~\cite{samsungPIM}, Alibaba's HB-PNM~\cite{alibabaPNM}, and SK hynix’s AiM~\cite{aim}, which promise substantial improvements in energy efficiency, performance, and bandwidth.
However, they also bring challenges, such as hardware design complexity, programming model design~\cite{tom-cnm}, and memory management~\cite{conda}. 

To address these challenges of CNM systems, several tools have been proposed in recent years. 
This includes simulators (e.g., PIMSimulator~\cite{samsungPIMSim}, ZSIM-Ramulator~\cite{zsim-ramulator}, uPIM~\cite{upim}) and programming frameworks~\cite{cinm, simplepim, to-pim-or-not, lin2023songc, samsungOneMCC}. 
While simulation tools are often technology-specific and slow, programming frameworks focus on abstracting away hardware details for easier programmability. 
They typically assume a fixed architecture, hindering detailed architectural exploration. 
Existing analytical models for CNM systems, and compilation frameworks that leverage them for offloading decisions~\cite{to-pim-or-not, tom-cnm}, often employ overly abstract hardware models, leading to inaccurate performance predictions (deviating by over an order of magnitude). 
Domain-specific frameworks like SADIMM~\cite{sadimm} deliver accurate estimates but only for sparse attention kernels.

In contrast, detailed cost models have been instrumental in CPU/GPU systems for decades, enabling design space exploration, optimization guidance, and intelligent offloading decisions~\cite{gpu_deep_cost, compoff}. For example,~\cite{compoff} proposes a neural network-based model for steering OpenMP offloading. Although CNM programming models share similarities with the offloading model for GPUs, their underlying memory-centric architectures, optimized for efficient memory access, differ significantly. Unlike GPUs with deep memory hierarchies, CNM systems typically have simpler computational units (e.g., RISC cores or domain-specific units) with direct access to larger memory pools. Memory management is often handled by custom logic~\cite{samsungPIM} or DMA engines~\cite{UPMEM}, but requires expert programming due to constraints such as data alignment in specific formats (e.g., transposed and interleaved in UPMEM). This, together with the access granularity, result in non-uniform latencies which makes the execution time estimation more challenging.  
Despite this added complexity, optimization and offloading responsibilities currently fall on the programmer, with the few existing optimizing compilers for CNM systems relying on simple, empirically-derived heuristics~\cite{cinm, xla-ndp, cimyin, imdp}.
These heuristics build atop established optimizations, that have proven efficient for multi-core CPUs and GPUs, like tiling for locality. 
We argue and show in this paper, that a proper modeling framework that captures the particularities of CNM architectures can reveal new optimization possibilities and thus better help guide the compilation process. 

\begin{figure}[t!]
\centering
\includegraphics[width=\columnwidth]{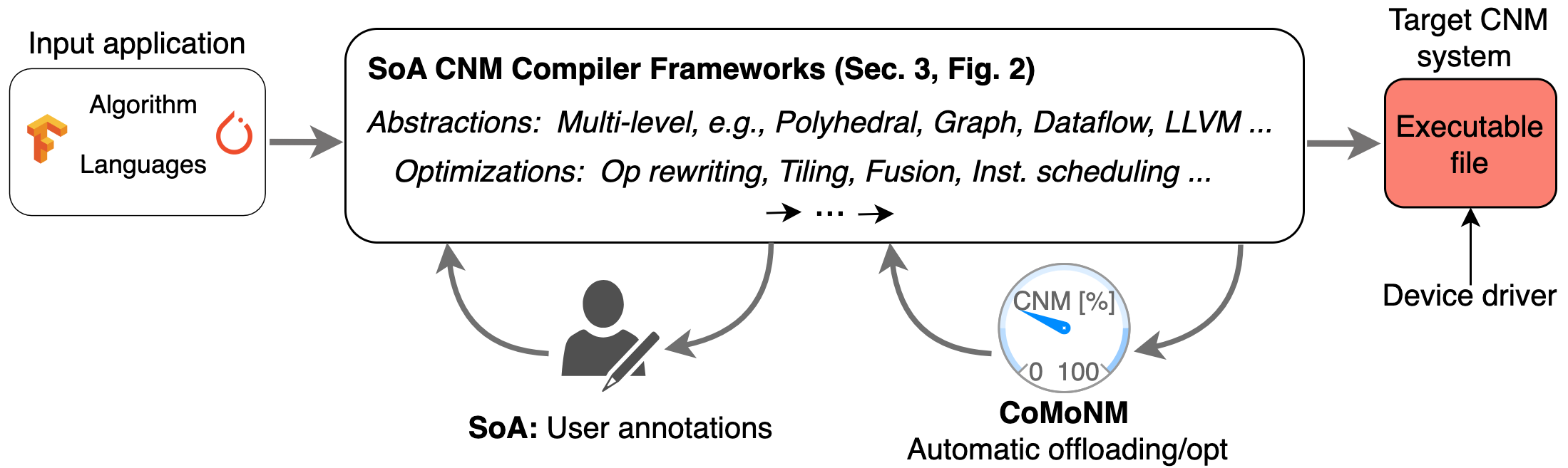}
\caption{\costmodelframework with SoA CNM compilation frameworks}
\label{fig:intro_fig}
\vspace{-5pt}
\end{figure}

Instead of \emph{importing} assumptions from mainstream architectures, accurate and fast cost models can facilitate a thorough understanding of the trade-offs in CNM architectures, empowering compiler, system, and performance engineers to make informed decisions. To this end, we introduce \textbf{\costmodelframework}, a generic cost modeling framework for CNM systems capable of estimating execution time in a few milliseconds. 

Fig.~\ref{fig:intro_fig} provides a high-level overview of  the abstractions and optimizations in state-of-the-art CNM compilation flows. 
These flows typically accept a high-level representation of the input application and progressively lower it to the target CNM system. Presently, user annotations, often based on educated intuitions, guide offloading code regions to CNM targets and applying various optimizations at different levels of abstraction.
These optimizations are done without sounds models of CNM systems. 
\costmodelframework offers different ways to interface to these frameworks (Fig.~\ref{fig:intro_fig}) to provide sound estimations of the effect of different transformations. 
As we show, this allows for better exploration of the design space. 
\costmodelframework takes input applications (at different abstraction levels for compatibility with existing frameworks), the target system's specifications, and a mapping of the application onto the system, and generates an estimated execution time for the given application and system configuration.

\costmodelframework's modular design and flexible front-end choices enable support for multiple CNM systems and should facilitate easy adaption to future architectures, as well as seamless integration into existing CNM compilation frameworks. We evaluate \costmodelframework using the open-source PriM benchmark suite~\cite{cnm-benches} and a set of machine learning applications. 
We show how the framework can be adapted to different CNM systems by providing a model and validating its  
accuracy against the UPMEM CNM system~\cite{UPMEM} and Samsung's HBM-PIM simulator. 
Applications can be directly represented using CNM-specific programming models, such as the UPMEM C interface~\cite{upmemSDK}, or through the several high-level domain specific languages (DSLs) that map to the linalg and affine dialects in MLIR~\cite{lattner2019mlir}. 
 These dialects are translated by \costmodelframework's front-end (\emph{cnm-gen}) into a virtual assembly we propose for CNM systems (\ourasm). 
 We demonstrate that integrating \costmodelframework into the state-of-the-art CINM compiler~\cite{cinm} can further enhance its performance and showcase its effectiveness in conducting design space exploration. %

Concretely, we make the following novel contributions. 
\begin{enumerate}
    \item We introduce \costmodelframework, a generic cost model 
    for CNM systems that rapidly and accurately estimates performance for a given application on a given CNM target system (Sec.~\ref{sec:uproto}). 
    \item We develop \emph{cnm-gen}, a front-end that takes a high-level hardware-agnostic representation of the input application and generates a virtual CNM assembly that is compatible with \costmodelframework (Sec.~\ref{subsec:assem-gen}).
    \item We propose a flexible mapping specification that captures the hierarchical nature of CNM systems, defining how computations are distributed across various levels (see Sec.~\ref{sec:mapping_detail}).
    \item We evaluate \costmodelframework on applications from diverse domains and validate its performance and accuracy against the the real UPMEM CNM system and the Samsung simulator (Sec.~\ref{sec:eval}). 
\end{enumerate}

We further demonstrate \costmodelframework's capability to rapidly explore huge design spaces, identifying non-intuitive transformations that outperform manual choices in state-of-the-art frameworks. Our evaluation on the UPMEM and Samsung CNM systems reveals that \costmodelframework's performance estimates are within 7.80\% of actual hardware measurements (UPMEM) and 2.99\% of simulator results (Samsung), exhibiting superior accuracy compared to UPMEM~\cite{upmemSDK} and uPIM~\cite{upim} simulators, while achieving 
seven orders of magnitude faster compared to the UPMEM and HBM-PIM simulators.

%% file: contents/new_background.tex
\section{CNM systems: Background}
This section provides background on CNM systems.
\label{sec:bg}
The core principle of CNM is to perform computations close to the memory by placing a processing element (PE) on or near the memory chip~\cite{cnm-review, khan2024landscape}. Due to architectural limitations of the memory system, these PEs are usually simple RISC cores are domain-specific custom logic units and the network within the memory across the PEs is either very limited or non-existent.
In the following, we describe two representative CNM systems, which we use to evaluate \costmodelframework.

\input{figures/motivation/compilers_opt_level}

\begin{figure*}[t!]
\begin{subfigure}[b]{0.15\textwidth}
    \begin{minipage}{\columnwidth}
\centering
\includegraphics[width=\columnwidth]{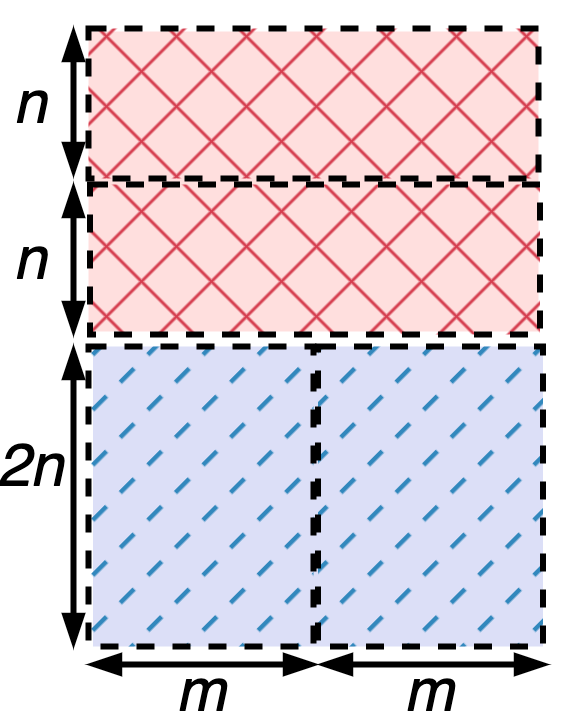}
    \end{minipage}%
\caption{An example of partitioned \matvec iteration space.}
\label{fig:upmem-matvec-iteration-space}
\end{subfigure}
\hfill
\begin{subfigure}[b]{0.235\textwidth}
\centering
\begin{minted}{c}
mram_read(addr_B, ...);
for(i = 0 ; i < I; i++){
 mram_read(addr_A, ...);
 ...
 c_C[i] = 0;
 for (j=0 ; j<J; j++){
  c_C[i]+=c_A[j]*c_B[j];
 }...}
mram_write(c_C,...);
\end{minted}
\caption{Tasklet code computing the iteration space with the dimensions $2n$ and $m$.}
\label{fig:upmem-matvec-single-mram}
\end{subfigure}
\hfill
\begin{subfigure}[b]{0.255\textwidth}
\centering
\begin{minted}{python}
for(i=0; i<I; i++){ ...
 for(j=0 ; j<J; j++){
   mram_read(addr_A, ...);
   mram_read(addr_B, ...);
   ...
   for(z=0 ; z<Z; z++){
    c_C[i]+=c_A[z]*c_B[z];
   }}}
mram_write(c_C, ...);
\end{minted}
\caption{Tasklet code computing the iteration space with the dimensions $n$ and $2m$.}
\label{fig:upmem-matvec-two-mrams}
\end{subfigure}
\hfill
\begin{subfigure}[b]{0.15\textwidth}
\centering
\begin{minted}{asm}
...
READ  E,R0,C
ADD   E,R1,C
WRITE E,R2,C
READ  O,R0,C
ADD   O,R1,C
WRITE O,R2,C
...
\end{minted}
\caption{Samsung's PE instructions - V1}
\label{fig:samsung-seq-1}
\end{subfigure}
\hfill
\begin{subfigure}[b]{0.15\textwidth}
\centering
\begin{minted}{asm}
...
READ  E,R0,C
READ  O,R0,C
ADD   E,R1,C
ADD   O,R1,C
WRITE E,R2,C
WRITE O,R2,C
...
\end{minted}
\caption{Samsung's PE instructions - V2}
\label{fig:samsung-seq-2}
\end{subfigure}
\caption{An example of possible partitioning iteration space of matrix-vector multiplication on tasklets.}
\label{fig:upmem-matvec-figs}
\end{figure*}

\subsection{The UPMEM's CNM system}
\label{upmem-background}
UPMEM is a CNM architecture featuring compute logic within the memory package but outside the memory arrays\cite{UPMEM}. In the latest UPMEM DIMMs, each DIMM consists of two ranks, each containing eight memory chips. Each memory chip includes eight data processing units (DPUs), each connected to a private 64 MB DRAM bank (referred to as main RAM or MRAM), accessible only by the corresponding DPU. In addition to the MRAM, each DPU has a private 64 KB SRAM-based scratchpad called working RAM (WRAM). The data inside the MRAM can be copied to the WRAM using the provided DMA engine inside the DPU. 
DPUs are simple RISC processors supporting only 32-bit integer operations, with other data type operations emulated in software. To hide DMA latency, each DPU features 24 hardware threads, known as tasklets, using an interleaved multi-threading (IMT) design, where only one tasklet can progress per cycle. The DPU execution pipeline consists of 14 stages, but due to dependency limitations, two consecutive instructions of a single tasklet must be 11 stages apart.

\subsection{The Samsung's HBM-PIM system}
Samsung's CNM system (HBM-PIM) integrates PEs within memory banks, optimized for executing machine learning applications. As described in~\cite{samsungPIM}, this design utilizes 2.5D high-bandwidth memory (HBM) DRAM. Each HBM-PIM instance comprises 16 memory channels, with each channel connected to a rank. Each rank consists of 16 banks, and a pair of banks is connected to a single PE. Each PE consists of 16 FPUs, each capable of performing multiplication and addition of 16-bit floating point operations. Additionally, each PE is also equipped with data registers (GRFs), control and instruction registers (CRF, SRF), and an internal control unit. The host processor manages PEs via standard DRAM interfaces.

%% file: figures/motivation/compilers_opt_level.tex
\begin{figure}[b]
\centering
\resizebox{\columnwidth}{!}{%
\begin{tikzpicture}[
  box/.style={draw=black, thick, rounded corners, minimum height=1cm, text centered, align=center, font=\small},
]

\draw[dashed, thick] (0,1) rectangle (11.5,5);
\draw[dashed] (3.83,1) -- (3.83,5);
\draw[dashed] (7.66,1) -- (7.66,5);

\node[box, fill=cyan!15, minimum width=11.225cm] at (5.755,4.2) {PUMA~\cite{puma}, PIMFlow~\cite{pimflow}, IMDP~\cite{imdp}};

\node[box, fill=green!15, minimum width=6.35cm] at (3.825,3) {CINM~\cite{cinm}, SongC~\cite{lin2023songc}, cim-mlc~\cite{cimmlc}, CIMFlow~\cite{cimyin}};

\node[box, fill=yellow!15, minimum width=3.5cm] at (1.9,1.8) {PIMCOMP~\cite{pimcomp}};
\node[box, fill=orange!15, minimum width=3.5cm] at (5.75,1.8) {TC-CIM~\cite{tc-cim}, OCC~\cite{occ},\\ Han et al.\cite{polyir_memrsitor}};
\node[box, fill=red!15, minimum width=3.5cm] at (9.6,1.8) {SIMDRAM~\cite{simdram}\\ Jin et al.~\cite{reram_compiler}};

\node[align=center, font=\small] at (1.915,0.5) {Operation graph};
\node[align=center, font=\small] at (5.745,0.5) {Loops};
\node[align=center, font=\small] at (9.575,0.5) {assembly};

\end{tikzpicture}
} %
\caption{Recent compiler frameworks for memory-centric accelerators, categorized by their optimization abstraction levels.}
\label{fig:compiler-abs-spec}

\end{figure}

%% file: contents/new_motiv.tex
\section{Related work and motivation}
\label{subsec:motiv}
To improve programmability and accessibility of memory-centric computing paradigms, several compiler frameworks have recently been introduced, including PIMFlow, PIMCOMP, SongC, and CINM \cite{occ, pimflow, simplepim, polyir_memrsitor, comprime, cinm, primo, xla-ndp, c4cam}.
Fig.~\ref{fig:compiler-abs-spec} provides an overview of these frameworks, classified based on the abstraction levels they leverage for optimization. As shown in the figure, optimizations include standard ones such as loop tiling, unrolling, and parallelization, as well as domain-specific strategies like weight replication in neural network layers, operation fusion, and pipelining. The choice and order of these optimizations significantly impact performance, making these decisions non-trivial. However, presently, in most frameworks, these decisions are either manual (left to the programmer) or semi-automated, relying on simplistic analytical models~\cite{to-pim-or-not} or inspired by optimization that had worked for muti-cores and GPUs. %

Simplified analytical models for CINM architectures often fail to capture the complexities of these systems, potentially leading to performance predictions that deviate by over an order of magnitude. This is due to the fact that these systems typically consist of multiple processing elements (PEs) with specialized memory access patterns, shared execution pipelines, and hardware components like DMA engines, which introduce unpredictable performance effects.
For the same reasons, manual optimization by programmers for these targets is also highly challenging.

\begin{figure}[b]

\centering
\input{figures/background/gemv_time}
\label{fig::gemv-motiv}
  \caption{Hardware execution vs. simulation times, for varying sizes of the matrix-vector multiplication kernel ($n$, matrix dimensions, $2^n \times 2^n$).}
\end{figure}

Consider a general matrix-vector multiplication(\matvec) kernel as an example. Fig.~\ref{fig:upmem-matvec-iteration-space} illustrates the computation space of this kernel, showing a possible tiling across four PEs (e.g., threads in UPMEM). While tile shapes impact data locality and communication as anticipated, other factors also significantly influence performance. For instance, the code in Fig.~\ref{fig:upmem-matvec-single-mram} and~\ref{fig:upmem-matvec-two-mrams} executes the tiled computation depicted in Fig.~\ref{fig:upmem-matvec-iteration-space}, with matching colors. In Fig.~\ref{fig:upmem-matvec-single-mram}, the tile shape and scratchpad size enable better data reuse for original vector, theoretically reducing data movement between WRAM and MRAM and improving performance. However, in practice, performance may be comparable to Fig.~\ref{fig:upmem-matvec-two-mrams} because, with fewer threads, DMA latency is fully amortized, leading to similar outcomes.

\begin{figure*}[t!]
  \centering
    \includegraphics[width=\textwidth]{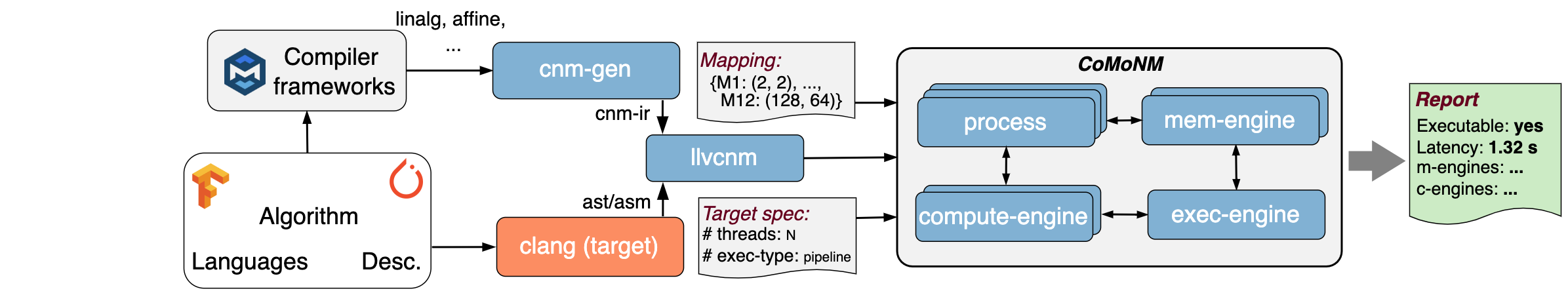}

  \caption{An overview of the \costmodelframework flow.}
  \label{fig::uproto_flow}
\vspace{-5pt}
\end{figure*}

Similarly, in Samsung’s HBM-PIM, performance is even more sensitive to low-level details. In HBM-PIM, the instruction order, type, bank type (even or odd), and even row and column addresses of consecutive instructions considerably impact the execution time. As an example, Fig.~\ref{fig:samsung-seq-1} and~\ref{fig:samsung-seq-2} present two instruction sequences performing the same task but yielding significantly different execution times ($\ge 2\times$).

Exploring such design spaces easily becomes computationally intractable. 
Even intelligently sampling the space can be prohibitively expensive due to 
manual rewriting and, specially, 
to the need for iterative testing involving real execution or simulations. 
This is illustrated in Fig.~\ref{fig::gemv-motiv} which shows the runtime of a single \texttt{gemv} configuration for  different sizes ($2^n \times 2^n$, where $n$ determines the matrix dimensions) on UPMEM’s real hardware (sys-ND), UPMEM’s simulator (sim-ND), the uPIM simulator (upim-ND) with different DIMM configurations (N), and Samsung’s HBM-PIM simulator. 
\costmodelframework addresses this challenge by providing fast (in milliseconds, independent of operand dimensions) and accurate ($\pm$5\% of real hardware) performance estimation for CNM targets. As an open-source tool, \costmodelframework is compiler-technology agnostic and can be integrated at the different levels of abstractions used in current CNMcompiler frameworks (see Fig.~\ref{fig:compiler-abs-spec}).

%% file: figures/background/gemv_time.tex
\pgfplotsset{compat = newest}
    \pgfplotsset{major grid style={dotted,aluminium2!50!black}}
    \begin{tikzpicture}
    \footnotesize
    \begin{axis}
    [
        width=\columnwidth,
         height=0.2\textwidth,
        ybar=0.5pt, %
        enlargelimits=0.15,
        enlarge y limits={upper, value=0.1},
        ylabel style={align=center},
        y label style={at={(-0.08,0.5)}},
        ylabel= Latency (ms),
        legend style={draw=none, fill=none},
        bar width=3pt,
        legend columns=4,
        ymode = log,
        log basis y={10},
        ymajorgrids=true,
        grid style=dashed,
        axis x line*=bottom,
        x tick label style={xshift=.0em, yshift=-.2em, rotate=0,anchor=east},
        yminorticks=true,
        legend style={at={(1.,1.45)},anchor=north east},
        symbolic x coords={13, 14, 15, 16, 17},
        xtick=data,
    ]
    
    \addplot+ [one_scheme_seven_colors] coordinates {
    (13, 10.394333)
    (14, 20.546333)
    (15, 57.730333)
    (16, 229.98)
    (17, 1928.59800)
    }; 

    \addplot+ [two_scheme_seven_colors] coordinates {
    (13, 10.432333)
    (14, 20.504333)
    (15, 40.666333)
    (16, 115.11)
    (17, 459.19)
    }; 
    
    \addplot+ [three_scheme_seven_colors] coordinates {
    (13, 172681.899)
    (14, 581244.587)
    (15, 2194845.013)
    (16, 8998347.093)
    (17, 35993388.37)
    }; 
       
    \addplot+ [four_scheme_seven_colors] coordinates {
    (13, 304337.578)
    (14, 1166427.478)
    (15, 4674455.552)
    (16, 16272538.97)
    (17, 65176771.24)
    }; 

    \addplot+ [five_scheme_seven_colors] coordinates {
    (13, 1987942.45)
    (14, 4859495.8662986755)
    (15, 12474123.0199337)
    (16, 33680132.15382099)
    (17, 94304370.03069878)
    }; 
       
    \addplot+ [six_scheme_seven_colors] coordinates {
    (13, 1958800.1251220703)
    (14, 5004326.936006546)
    (15, 11595248.496055603)
    (16, 31307170.939350132)
    (17, 87660078.63018036)
    }; 

    \addplot+ [seven_scheme_seven_colors] coordinates {
    (13, 16081)
    (14, 59604)
    (15, 46132)
    (16, 183338)
    (17, 733352)
    }; 
    
    \legend{sys-8D, sys-16D, sim-8D, sim-16D, upim-8D, upim-16D, hbm-pim}
    \end{axis}
\end{tikzpicture}
\caption{\matvec execution time for different configurations}

%% file: contents/costmodelInputs.tex
\section{\costmodelframework: A Cost model for CNM Systems}
\label{sec:uproto}

This section describes the different components and functionality of \costmodelframework's flow, as illustrated in Fig.~\ref{fig::uproto_flow}.

\subsection{Overview}
\label{subsec:overview}

The \costmodelframework takes an input application, a target CNM system specification, and a user-defined mapping that outlines how the application is mapped onto the CNM system. As output, it produces the estimated execution time of the application on the given system. The input application can be written in our target-agnostic virtual assembly (\ourasm, see Fig.~\ref{fig::uproto_flow}), a high-level language like C/C++, or a domain-specific language such as \linalg, enabling support for various ML frameworks (via their existing front-ends to produce \linalg) (see Sec.~\ref{subsec:app-rep}). The \emph{cnm-gen} module translates these high-level representations into the virtual assembly compatible with \costmodelframework (see Sec.~\ref{sec:mapping_detail}). The mapping specification, represented as a vector of tuples, captures the CNM system's hierarchical organization (see Sec.~\ref{sec:mapping_detail}). The target system description is a configuration file that defines the target CNM system (UPMEM, HBM-PIM) and its parameters.

\costmodelframework models main components of the CNM system, including it \emph{compute units} (e.g., pipelines, ALUs), the \emph{memory engine} (capturing different levels of the memory hierarchy including private and shared memories), and the \emph{process} being executed on the execution units (e.g, threads in UPMEM), as shown in Fig.~\ref{fig::uproto_flow}. A process refers to the smallest unit of a kernel being executed on the system (e.g., threads in UPMEM or kernel instances run by a PIMUnit in HBM-PIM).
Multiple processes can share an compute-engine and a mem-engine. For example, in UPMEM, up to 24 threads share the same execution pipeline, MRAM, and WRAM within a DPU. Similarly, in HBM-PIM, a PIMUnit runs 16 parallel instances of the same kernel, where all instances share register files and access the same two memory banks.
The \emph{exec-engine} module in \costmodelframework stores the base latencies of different \ourasm instructions, extracted through rigorous system profiling under various conditions. This information is used by both the \emph{compute-engine} and \emph{mem-engine} modules to simulate latency accurately. %

In the following, we explain each of these components and their implementation in detail. 

\subsection{Input application representation}
\label{subsec:app-rep}

CNM systems typically support high-level languages (e.g., C++ for UPMEM, various ML frameworks for HBM-PIM) and provide proprietary toolchains to translate these high-level application representations into target-specific instructions. However, neither these high-level representations nor their resulting intermediate representations are adequately abstract for our \costmodelframework, which requires generalization across multiple CNM systems. To address this, we introduce a new virtual assembly representation that abstracts over the ISAs of various CNM targets (see Sec.~\ref{subsec:assem-gen}). While applications can be directly written in this virtual assembly, this process can be tedious and time-consuming.

Alternatively, using higher-level abstractions like those in MLIR enables broader support for diverse input languages and frameworks. For example, MLIR’s \linalg abstraction~\cite{mlir} is a well-established backend for numerous ML frameworks and DSLs. We develop necessary conversions from \linalg to our virtual assembly (explained in Sec.~\ref{subsec:assem-gen}), allowing \costmodelframework to seamlessly integrate with any framework or language that targets \linalg. This not only ensures compatibility with a wide range of application representations but, more importantly, enables the integration of \costmodelframework into CNM compiler frameworks, as discussed in Sec.~\ref{subsec:motiv}, since the input to these frameworks is often represented in \linalg.

\begin{figure}[tbh]
    \centering
\begin{lstlisting}[numberstyle=\scriptsize\color{codegray},language=mlir,basicstyle=\scriptsize\ttfamily,linerange={1-20},breaklines=true,postbreak=\mbox{\textcolor{red}{$\hookrightarrow$}\space},frame=tb,linewidth=\columnwidth]
#m1 = affine_map<(d0,d1)->(d0,d1)>
#m2 = affine_map<(d0,d1)->(d1)>
#m3 = affine_map<(d0,d1)->(d0)>
linalg.generic{indexing_maps=
  [#m1,#m2,#m3], iterator_types=["parallel","reduction"]}
  ins(%a0, %a1 : memref<MxNxi32>,memref<Nxi32>) 
  outs(%a2: memref<Mxi32>){
 ^bb0(%in:i32, %in_0:i32, %out:i32):
   %0 = arith.muli %in, %in_0 :i32
   %1 = arith.addi %0, %out :i32
   linalg.yield %1 :i32}
---------------------------------------------
def forward(self, input: Tensor, ...) -> ...:
    ...
    matmul = torch.matmul(input, (others))
    ...
    return output

\end{lstlisting}
    \caption{\matvec kernel in different DSLs.}
    \label{fig:matvec-dsl}
\vspace{-5pt}
\end{figure}

Figs.~\ref{fig:upmem-matvec-single-mram} and~\ref{fig:upmem-matvec-two-mrams} present code snippets of the main loop nest for the \matvec kernel written in UPMEM’s C language. The same can also be expressed in other representations, such as \linalg and \texttt{TorchScript} (see Fig.~\ref{fig:matvec-dsl}), all of which are supported by \costmodelframework. \costmodelframework’s conversion process from \linalg to virtual assembly involves analyzing the \linalg IR to extract critical information, including \emph{loop ranges}, \emph{indexing maps}, \emph{affine maps}, \emph{memory access patterns} for inputs and outputs, and \emph{dependency details}. This information, combined with target-specific mapping details (explained in the next section), is utilized by the \texttt{cnm-gen} module to generate \costmodelframework-compatible virtual assembly code (e.g., Fig.~\ref{asm:mav_simple_asm}).

\subsection{Mapping specification}
\label{sec:mapping_detail}
\costmodelframework takes a user-defined mapping as input, which explicitly specifies how the kernel's iteration space is partitioned across the compute units at each level of the system's hierarchy. For instance, in the context of UPMEM, this involves defining parameters such as the number of ranks, the number of DPUs per rank, the number of threads per DPU, and the workload distribution strategy at each hierarchical level. To represent these partitioning choices, we introduce a generic mapping representation that captures the iteration space partitioning strategy across the system's compute hierarchy. Making the mapping an explicit input to \costmodelframework enables integration with auto-tuning frameworks for design space exploration as well as facilitating effective exploration of transformation spaces within CNM compiler frameworks.

\subsubsection{Mapping iteration space to a CNM target} 
\begin{figure}[tbh]
\begin{lstlisting}[numberstyle=\footnotesize\color{codegray},language=mlir,basicstyle=\scriptsize\ttfamily,linerange={1-20},breaklines=true,postbreak=\mbox{\textcolor{red}{$\hookrightarrow$}\space},frame=tb,linewidth=\columnwidth]
Upmem_map: [
    M1: {(2, 9), (64, 1), (2, 4), (2, 32)},   (a)
]
------------------------------------------------
Upmem_map: [
    M2: {(4, 1), (8, 8 ), (1, 4), (16, 20)},   
    M3: {(1, 4), (2, 4 ), (8, 1), (12, 32)},  (b)
    M1: {(4, 4), (5, 5 ), (1, 1), (16, 16)},
    M4: {(1, 1), (5, 16), (16,1), (4 , 12)}
]
------------------------------------------------
Hbmpim_map: [
    M1: {(4, 4), (2, 8), (8, 2), (4, 9)},
    M3: {(4, 4), (2, 8), (8, 2), (4, 9)},     (c)
    M4: {(4, 4), (2, 8), (8, 2), (4, 9)},
    M2: {(4, 4), (2, 8), (8, 2), (4, 9)},
]

\end{lstlisting}
    \caption{Two \matvec mappings for UPMEM (a and b) and a mapping for HBM-PIM (c).}
    \label{fig:matvec-map}
\end{figure}

To map an $N$-dimensional iteration space onto a target system with $K$ levels of compute-units hierarchy, we define a map $m$ as:  
\begin{equation}
m= \{[d^{1}_{1}, d^{1}_{2}, ..., d^{1}_{N}], ..., [d^{K+1}_{1}, d^{K+1}_{2}, ..., d^{K+1}_{N}]\} 
\end{equation}
This map is a vector of $K+1$ tuples, each containing $N$ elements. An element in the \textit{i}-th tuple at the \textit{j}-th position represents the range of iterations for the \textit{i}-th level of the hierarchy along the \textit{j}-th dimension, i.e., how many computations along a given dimension are performed at a particular level of the system hierarchy.

\begin{figure}[tbh]
\centering
\begin{subfigure}[b]{\columnwidth}
\centering
\includegraphics[width=\textwidth]{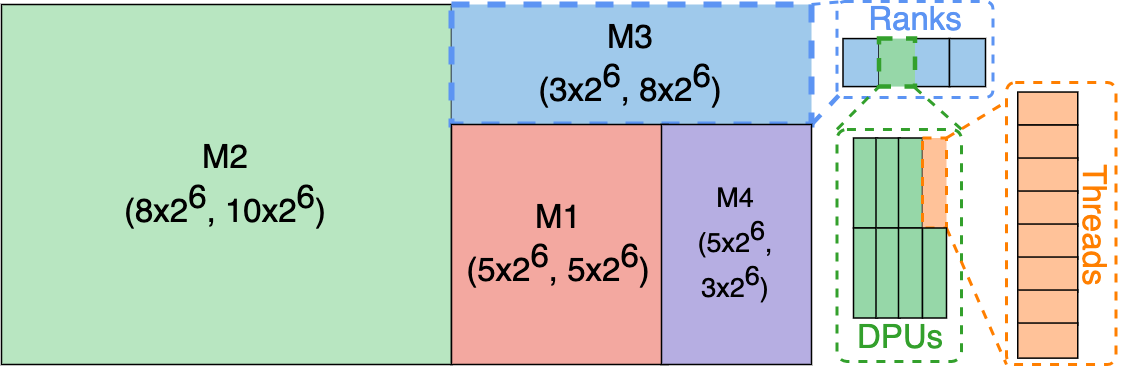}
\caption{Graphical mapping representation for UPMEM.}
\label{fig:mv-upmem-map-graph}
\end{subfigure}
\begin{subfigure}[b]{\columnwidth}
\centering
\includegraphics[width=\textwidth]{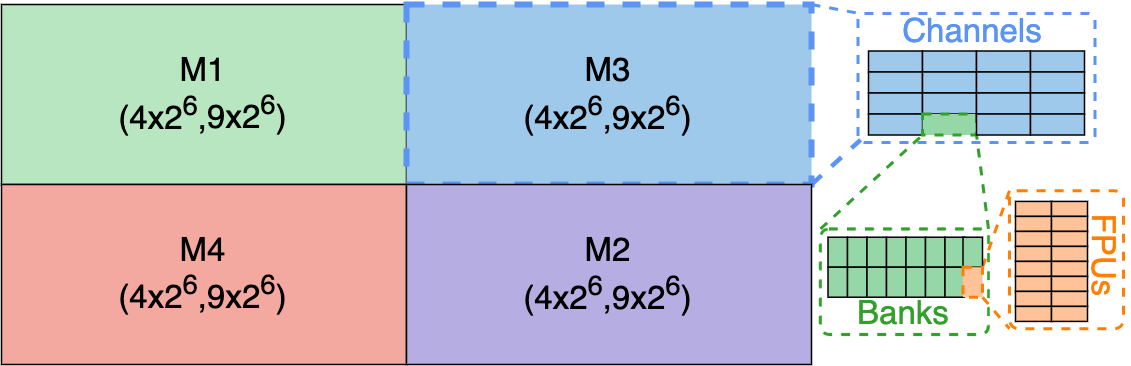}
\caption{Graphical mapping representation for HBM-PIM}
\label{fig:mv-HBM-PIM-map-graph}
\end{subfigure}
\caption{An example of partitioning iteration space of \matvec kernel for different CNM systems.}
\label{fig:matvec-graph-figs}
\end{figure}

The iteration space and dependencies of the \matvec kernel depicted in Fig.~\ref{fig:matvec-dsl} are defined by the loop nest structure and the iteration range at each level. 
With iteration space dimensions of $(8 \times 2^6,\ 18 \times 2^6)$, a potential mapping vector that uniformly distributes computation across compute units is illustrated in Fig.~\ref{fig:matvec-map}a. However, in practice, executing the entire kernel on compute units might not be feasible or desirable due to memory constraints in CNM systems, necessitating the partitioning of the iteration space. \costmodelframework supports partitioning the iteration space into multiple subspaces, as shown in Figure~\ref{fig:matvec-graph-figs}, defining a seperate mapping for each subspace, and mapping it independently to the target system, as demonstrated in Figs.~\ref{fig:matvec-map}b and~\ref{fig:matvec-map}c. This fine-grained handling of individual tiles also enables the exploration of the impact of different tile shapes and sizes on the overall execution time.

For UPMEM, the hierarchy levels consist of ranks, DPUs, and threads (tasklets), whereas for HBM-PIM, they include channels, banks, and FPUs. For example, in UPMEM’s M3 mapping (see Figure~\ref{fig:matvec-map}b), iterations are distributed across $1 \times 4 = 4$ ranks, $2 \times 4 = 8$ DPUs per rank, $8 \times 1 = 8$ threads per DPU, and a $12 \times 32$ iteration space per thread. Generally, each tuple at a hierarchy level specifies the division factor for the iteration space dimensions. In the M3 mapping for UPMEM, the tuple $(1, 4)$ at the first level indicates that the iteration space is partitioned into four subspaces along the second dimension. Similarly, at the DPU level, the tuple $(2, 4)$ signifies that each rank’s assigned space is divided two times along the first dimension and four times along the second, resulting in eight subspaces. This partitioning logic applies consistently across all hierarchy levels.

\subsubsection{Order and dependencies}
Tiling the iteration space can introduce dependencies depending on the application. For example, in UPMEM's M3 mapping (Figure~\ref{fig:matvec-map}), which uses four ranks, partitioning along the second dimension creates a dependency (reduction). Conversely, partitioning along the second dimension (i.e., using a mapping of (4, 1)) would not introduce such a dependency. If the target architecture supports accumulation of partial results at the rank level, it is performed there; otherwise, partial results are sent to the higher level in the hierarchy for accumulation (CPU in the case of UPMEM).

Dependencies can also arise across different mappings. For instance, UPMEM’s M2 and M3 mappings and HBM-PIM’s M1 and M3 mappings (Fig.~\ref{fig:matvec-map}) partition the iteration space along a dimension with dependencies, requiring eventual accumulation of partial results from these subspaces. Although each subspace can start computation independently, their partial results must be accumulated if they share a dependency dimension, thus impacting 
the overall execution time. %

To manage the allocation and execution of the different maps, our mapping specification uses indices to denote execution precedence. In both UPMEM and HBM-PIM mappings, instances of M1, M2, M3, and M4 are allocated and executed sequentially. Allocation proceeds along the first dimension and advances in order, as shown in the example. 
By only taking the mapping specification and input application, \costmodelframework automatically and faithfully models all these heterogeneous tiling, allocation, dependencies, and partial results accumulation scenarios and accurately estimates their latencies. %

\subsection{CNM virtual assembly code generation}
\label{subsec:assem-gen}
\begin{figure}[t!]
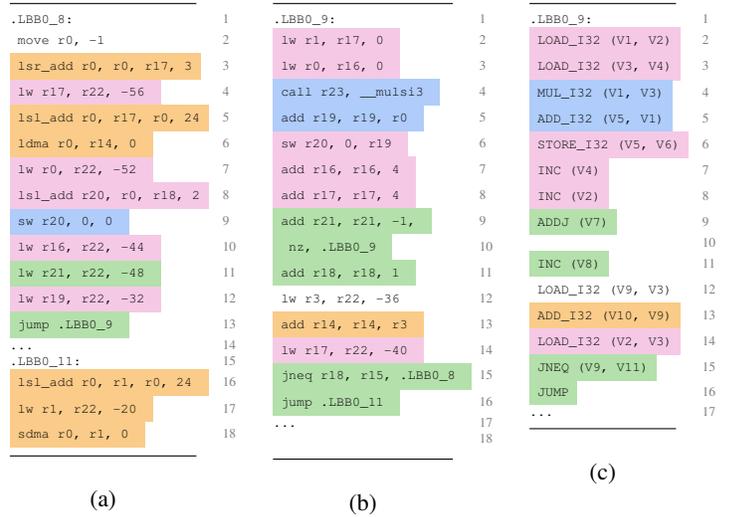

\centering
\begin{subfigure}[t]{0.28\columnwidth}
\lstset{escapeinside={<@}{@>}}
\begin{lstlisting}[numberstyle=\tiny\color{codegray},numbers=right,language={[x86masm]Assembler},basicstyle=\tiny\ttfamily,linerange={14-31},breaklines=true,postbreak=\mbox{\textcolor{red}{$\hookrightarrow$}\space},frame=tb,linewidth=\columnwidth]
.LBB0_8:                               
<@\colorbox{minted_ins_norm}{move r0, -1}@>
<@\colorbox{minted_ins_mram}{lsr\_add r0, r0,  r17, 3\label{asm_code:addr1}}@>
<@\colorbox{minted_ins_wram}{lw r17, r22, -56       }@>
<@\colorbox{minted_ins_mram}{lsl\_add r0, r17, r0, 24\label{asm_code:addr2}}@>
<@\colorbox{minted_ins_mram}{ldma r0, r14, 0        }@>
<@\colorbox{minted_ins_wram}{lw r0, r22, -52        \label{asm_code:addr5}}@>
<@\colorbox{minted_ins_wram}{lsl\_add r20, r0, r18, 2\label{asm_code:addr6}}@>
<@\colorbox{minted_ins_comp}{sw r20, 0, 0           \label{asm_code:oreset}}@>
<@\colorbox{minted_ins_wram}{lw r16, r22, -44       \label{asm_code:addr7}}@>
<@\colorbox{minted_ins_cf}{lw r21, r22, -48       \label{asm_code:reset_ind}}@>
<@\colorbox{minted_ins_wram}{lw r19, r22, -32       \label{asm_code:accres}}@>
<@\colorbox{minted_ins_cf}{jump .LBB0\_9           }@>
...
.LBB0_11:
<@\colorbox{minted_ins_mram}{lsl\_add r0, r1, r0, 24 \label{asm_code:addr3}}@>
<@\colorbox{minted_ins_mram}{lw r1, r22, -20        \label{asm_code:addr4}}@>
<@\colorbox{minted_ins_mram}{sdma r0, r1, 0         }@>
\end{lstlisting}
\caption{}
\label{asm:matvec-level1}
\end{subfigure}
\hfill
\begin{subfigure}[t]{0.27\columnwidth}
\lstset{escapeinside={<@}{@>}}
\begin{lstlisting}[numberstyle=\tiny\color{codegray},numbers=right,language={[x86masm]Assembler},basicstyle=\tiny\ttfamily,linerange={16-33},breaklines=true,postbreak=\mbox{\textcolor{red}{$\hookrightarrow$}\space},frame=tb,linewidth=\columnwidth]
.LBB0_9:                                
<@\colorbox{minted_ins_wram}{lw r1, r17, 0          \label{asm_code:load1}}@> 
<@\colorbox{minted_ins_wram}{lw r0, r16, 0          \label{asm_code:load2}}@>
<@\colorbox{minted_ins_comp}{call r23, \_\_mulsi3     \label{asm_code:mult}}@> 
<@\colorbox{minted_ins_comp}{add r19, r19, r0       \label{asm_code:acc}}@>
<@\colorbox{minted_ins_wram}{sw r20, 0, r19         \label{asm_code:store}}@>
<@\colorbox{minted_ins_wram}{add r16, r16, 4        \label{asm_code:inc1}}@>
<@\colorbox{minted_ins_wram}{add r17, r17, 4        \label{asm_code:inc2}}@>
<@\colorbox{minted_ins_cf}{add r21, r21, -1,      \label{asm_code:jump}}@>
<@\colorbox{minted_ins_cf}{    nz, .LBB0\_9        }@>
<@\colorbox{minted_ins_cf}{add r18, r18, 1        \label{asm_code:outacc}}@>
<@\colorbox{minted_ins_norm}{lw r3, r22, -36}@>
<@\colorbox{minted_ins_mram}{add r14, r14, r3       }@>
<@\colorbox{minted_ins_wram}{lw r17, r22, -40       }@>
<@\colorbox{minted_ins_cf}{jneq r18, r15, .LBB0\_8 \label{asm_code:outj}}@>
<@\colorbox{minted_ins_cf}{jump .LBB0\_11          }@>
...
<@\colorbox{minted_ins_norm}{}@>
\end{lstlisting}
\caption{}
\label{asm:matvec-level2}
\end{subfigure}
\hfill
\begin{subfigure}[t]{0.22\columnwidth}
\lstset{escapeinside={<@}{@>}}
\begin{lstlisting}[numberstyle=\tiny\color{codegray},numbers=right,language={[x86masm]Assembler},basicstyle=\tiny\ttfamily,linerange={1-18},breaklines=true,postbreak=\mbox{\textcolor{red}{$\hookrightarrow$}\space},frame=tb,linewidth=\columnwidth]
.LBB0_9:                                
<@\colorbox{minted_ins_wram}{LOAD\_I32 (V1, V2)}@> 
<@\colorbox{minted_ins_wram}{LOAD\_I32 (V3, V4)}@>
<@\colorbox{minted_ins_comp}{MUL\_I32 (V1, V3)}@> 
<@\colorbox{minted_ins_comp}{ADD\_I32 (V5, V1)}@>
<@\colorbox{minted_ins_wram}{STORE\_I32 (V5, V6)}@>
<@\colorbox{minted_ins_wram}{INC (V4)}@>
<@\colorbox{minted_ins_wram}{INC (V2)}@>
<@\colorbox{minted_ins_cf}{ADDJ (V7)}@>

<@\colorbox{minted_ins_cf}{INC (V8)}@>
<@\colorbox{minted_ins_norm}{LOAD\_I32 (V9, V3)}@> 
<@\colorbox{minted_ins_mram}{ADD\_I32 (V10, V9)}@>
<@\colorbox{minted_ins_wram}{LOAD\_I32 (V2, V3)}@> 
<@\colorbox{minted_ins_cf}{JNEQ (V9, V11)}@>
<@\colorbox{minted_ins_cf}{JUMP}@>
...
\end{lstlisting}
\caption{}
\label{asm:mav_simple_asm}
\end{subfigure}
\caption{Assembly code of innermost loop (a) and outermost loop (b) of a \matvec Kernel, and outermost loop in \ourasm (c).}
\label{asm:matvec}
\vspace{-5pt}
\end{figure}

Fig.~\ref{asm:matvec-level1} and Fig.~\ref{asm:matvec-level2} show parts of the optimized assembly code for the \matvec kernel generated with the \texttt{-O3} optimization level using UPMEM’s compiler (\texttt{dpu-clang}). Fig.~\ref{asm:matvec-level1} highlights a snippet of the outermost loop (\texttt{LBB0\_8}), while Fig.~\ref{asm:matvec-level2} illustrates the innermost loop (\texttt{LBB0\_9}). 
These instructions are grouped based on their interaction with the UPMEM's DPU components: Orange lines represent instructions that interact with the DMA to load or store data to MRAM or compute MRAM memory addresses. Pink lines show instructions that transfer data between the arithmetic units and WRAM or calculate WRAM memory addresses. Blue lines indicate arithmetic operations or result initialization performed on the data.

\costmodelframework's target-code parser processes UPMEM’s assembly and translates it into our virtual assembly representation called \ourasm, which is supported by the framework. 
Fig.~\ref{asm:mav_simple_asm} illustrates the innermost loop of the \matvec kernel in \ourasm. 
The representation consists of basic instructions such as \texttt{LOAD}, \texttt{STORE}, \texttt{MUL}, etc., for different data types. The instruction operands are virtual and are not used for the functional correctness of the kernel. 
In UPMEM’s case, the operands are not taken into account for latency estimation. However, in HBM-PIM, they are considered, as instruction latency depends on the data's physical location.

\begin{figure}[b]
\begin{lstlisting}[numberstyle=\scriptsize\color{codegray},numbers=right,language=mlir,basicstyle=\scriptsize\ttfamily,linerange={1-18},breaklines=true,postbreak=\mbox{\textcolor{red}{$\hookrightarrow$}\space},frame=tb,linewidth=\columnwidth]
in1 = Operand(A, TENSOR, DTYPE)
in2 = Operand(B, TENSOR, DTYPE)
o1 = Output(C, TENSOR, DTYPE)
induction_var1 = Index(I, M)
...
in1_L2_READ = MEM_OP(in1, L1, LOAD)
in1_L2_READ.setIndVar(induction_var_1)
...
MulOp = Operation(ADD, [in1, in2], t1, DTYPE)
ADDOp = Operation(ADD, [t1, t2], t2, DTYPE)
...
loop1 = For([A_L2Read, B_L2Read, A_L2_ADDR_INC,
    B_ADDR_INC, MulOp, ADDOp, C_L2Write],
    induction_var2)
loop2 = For([A_L3Read, loop1, A_L3_ADDR_INC,
    Res_INC], induction_var1)
ret ([B_L3_Read, loop2, C_L3_Write])
\end{lstlisting}
\caption{\texttt{CNM} IR for MV.}
\label{fig:cnm-ir}
\end{figure}

The \texttt{linalg} input to \costmodelframework is progressively lowered into \ourasm in two stages. First, the high-level \texttt{linalg} IR is transformed into a custom \texttt{CNM} IR, which facilitates loop representation and simplifies conversion to the low-level \ourasm. For our \matvec example, Fig.~\ref{fig:cnm-ir} illustrates the \texttt{CNM} IR generated from the \texttt{linalg} code in Fig.~\ref{fig:matvec-dsl}. The \texttt{CNM} IR captures \textit{operand structure and type} (e.g., array or scalar, int, float, double), \textit{loop induction variables} (line 4), \textit{memory requests} (including request type, size, and address), \textit{CNM operations} (line 9), \textit{loops} (lines 12–17), and the \textit{execution order}, which specifies a sequential list of operations (line 18).

Next, the \texttt{CNM} IR is translated into \ourasm, with the conversion tailored to the specification of the target
System-specific characteristics, such as the number of memory hierarchy levels, influence the generation of \texttt{LOAD} and \texttt{STORE} instructions and control flow. 
For example, in UPMEM DPUs, there are three memory levels—MRAM, WRAM, and thread register files—whereas in HBM-PIM, there are only two: register files and banks.%

%% file: contents/costmodel_framwork.tex
\section{System modeling and performance estimation}
This section details the primary components of \costmodelframework used to model a CNM system and describes its execution engine for performance estimation.

\subsection{Modeling a CNM system}
As illustrated in Fig.~\ref{fig::uproto_flow}, \costmodelframework is composed of three primary components that model a CNM system.
\subsubsection{Compute-engine} This models the compute subsystem responsible for executing instructions. For instance, in the case of UPMEM, it models the behavior of the DPU execution pipeline, while for HBM-PIM, it models the operations of the FPUs inside the PIMUnits. More specifically, the \emph{compute-engine} implements two different modes: \texttt{PIPELINE} and \texttt{VECTOR\_PROC}. During initialization, this module takes the target system specification as input to configure accordingly. Additional constraints can be incorporated by overriding the base \texttt{accept} function. For example, in the case of UPMEM, the execution model can enforce that two instructions from the same tasklet must be separated by at least 11 pipeline stages.

\subsubsection{Mem-engine}
The \emph{mem-engine} models different levels of memory hierarchy %
e.g., MRAM and WRAM in UPMEM, registers and banks in HBM-PIM). It captures the different memory access behaviors, including the access restrictions and possible contention between the processes. It supports different types of memory access mechanisms, such as: \textit{1. Blocking access}, like the bank interfaces in HBM-PIM that move data between the register file and memory banks, and \textit{2. Non-blocking access}, such as the use of DMA engines in UPMEM to transfer data between WRAM and MRAM.

\input{figures/overview/new_new_latency_calc_algo}

\subsection{\costmodelframework Execution engine}
\label{subsec:uproto-detailed}
The \costmodelframework execution engine takes the modeled CNM system (as detailed in the preceding section) and the input application (in \ourasm format), and estimates performance. For latency numbers, it relies on pre-stored look-up tables (LUTs), derived from a thorough performance characterization of the target CNM systems.

Algorithm~\ref{alg::detailed_sim} provides pseudocode for how the \costmodelframework estimates application performance. Lines \ref{algo:line:init_start}-\ref{algo:line:init_end} initializes the variables for the prediction. The core body of the compute engine (Lines \ref{algo:line:perf_start}-\ref{algo:line:perf_end}), extracts representative instances/slices from the iteration space using \texttt{get\_portion()} (Lines \ref{algo:line:get_portion1} and \ref{algo:line:get_portion2}). These instances are simulated in an instruction-accurate fashion via the \texttt{simulate} function, and the observed trends from these slices are extrapolated to estimate performance across the entire iteration space (Line \ref{algo:line:extrapolate}). The \texttt{get\_portion(space, 1)} function returns the smallest subspace that executes all of the instructions at least once, with Lines 11 and 12 extracting two such subspaces. 

Based on the CNM system and its compute units — either a shared pipeline across threads (UPMEM) or a vector-processor (HBM-PIM) (Lines ~\ref{algo:line:pipeline} and ~\ref{algo:line:vector_proc}) — the \texttt{simulate} function  dispatches instructions to the compute-engine or mem-engine, depending on the instruction type (e.g., READ vs. ADD). Once dispatched, the corresponding unit is marked as blocked until execution completes, accurately reflecting system behavior in different scenarios. For instance, in UPMEM systems, kernels with DMA operations executed across multiple threads issuing concurrent DMA requests have varying execution times due to shared resource contention.

\begin{figure}[t!]
\centering
\includegraphics[width=\columnwidth]{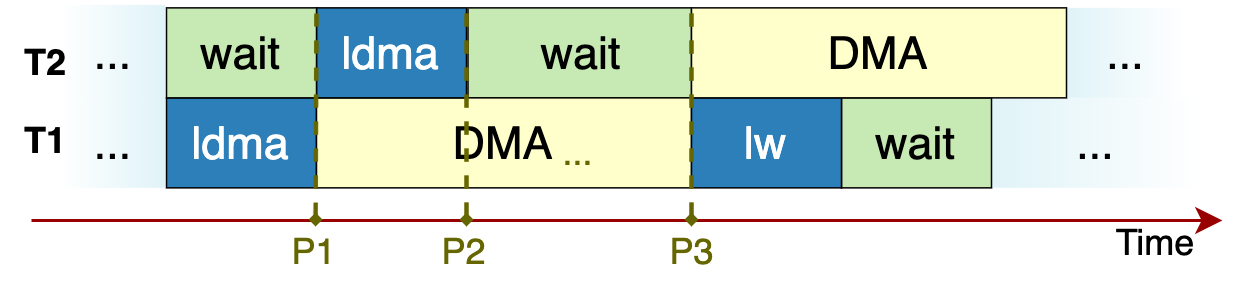}
\caption{Thread contention during DMA access.}
\label{fig:mv-flow}
\vspace{-7pt}
\end{figure}

Fig.~\ref{fig:mv-flow} illustrates this scenario, with the X-axis representing time and Y-axis showing different threads. When thread (T1 and T2) issue DMA instructions simultaneously, only one request can enter the DMA pipeline at a time. In this case, T2 stalls until T1's DMA request is issued (point p1). After T1 enters the DMA engine, T2 resumes pipeline execution but must wait again for the DMA to become available (point p3). These blocking behaviors cause irregular delays. However, as a single DMA operation has a fixed latency and kernel behaviors are usually repetitive, \costmodelframework captures patterns from a few executed instances to estimate overall performance.

\vspace{-7pt}

%% file: figures/overview/new_new_latency_calc_algo.tex
\begin{algorithm}[h]
\footnotesize
\caption{\costmodelframework execution engine}
\label{alg::detailed_sim}
\textbf{Inputs: } $ce \gets $ compute\_engine, $me \gets $ mem\_engine, kernel, $m \gets $ mapping \\
\textbf{Output: } ET \textcolor{blue}{// ET: Estimated execution time} \\
mode $\gets$ $ce$.mode \textcolor{blue}{// target specific}\label{algo:line:init_start}\\
ts $\gets$ [\texttt{Free}, $\ldots$, \texttt{Free}],\quad t\_pc $\gets$ [0, $\ldots$, 0] \textcolor{blue}{// ts: threads state}\\
vec\_state $\gets$ Free\\
$is \gets m$.iterationSpace\label{algo:line:init_end}\\

PE $\gets$ perf\_estimate ($ce$, kernel, $is$)\label{algo:line:perf_start}\\

\textbf{function} perf\_estimate ($ce$, kernel, $is$)\\
step1 $\gets -1$, step2 $\gets 1$, space\_coef $\gets 1$\\
space1 $\gets$ \texttt{inf},\quad space2 $\gets$ \texttt{empty}\\

\While{$|2 - \frac{step2}{step1}| > \alpha$ \KwSty{and} $space1 \neq space2$}{
    space1 $\gets$ \texttt{get\_portion}($is$, space\_coef)\label{algo:line:get_portion1}\\
    space2 $\gets$ \texttt{get\_portion}($is$, space\_coef $\times$ 2)\label{algo:line:get_portion2}\\
    step1 $\gets$ simulate(kernel, $ce$, space1)\\
    step2 $\gets$ simulate(kernel, $ce$, space2)\\
    space\_coef $\gets$ space\_coef + 1
}
coef $\gets$ \texttt{remaining\_space}($is$)\\
\Return step2 $\times$ coef \label{algo:line:extrapolate}

\textbf{function} simulate(kernel, $ce$, $is$)\label{algo:line:flat_ins}\\
ins\_trace $\gets$ get\_flat\_ins(kernel, $is$) \\
pc $\gets$ 0\\
\While{pc $\neq$ len(ins\_trace)}{
    \If{$ce$.mode == \texttt{PIPELINE}}{
    \label{algo:line:pipeline}
        \ForEach{tid in threads}{
            \If{$ts$[tid] == \texttt{FREE}}{
                ins $\gets$ kernel.get\_ins(t\_pc[tid])\\
                \If{ce.accept(ins, tid) \KwSty{or} me.accept(ins, tid)}{
                    ts[tid] $\gets$ \texttt{BLOCKED}\\
                    t\_pc[tid] $\gets$ t\_pc[tid] + 1\\
                    \textbf{break}
                }
            }
        }
        pc $\gets$ min(t\_pc)
    }
    \ElseIf{$ce$.mode == \texttt{VECTOR\_PROC}}{
    \label{algo:line:vector_proc}
        ins $\gets$ kernel.get\_ins(vector\_pc)\\
        \If{ce.accept(ins) \KwSty{or} me.accept(ins)}{
            vec\_state $\gets$ \texttt{BLOCKED}\\
            pc[tid] $\gets$ pc[tid] + 1\\
        }
    }
    cycle\_count $\gets$ cycle\_count + 1
}
\Return cycle\_count \label{algo:line:perf_end}
\end{algorithm}

%% file: contents/evaluation.tex
\section{Evaluation}
\label{sec:eval}
This section explains our experimental setup, benchmarks and the evaluation results.  
\subsection{Experimental setup and configurations}
\label{subsec:setup}
We evaluate the modeling and prediction capabilities of \costmodelframework using UPMEM (\emph{upmem}) and HBM-PIM (hbmpim) CNM systems. For UPMEM, we use the clang-generated assembly (\emph{c}) as well as the assembly instructions generated by \costmodelframework from \texttt{linalg} (\emph{gen}). For HBM-PIM, we rely solely on the \texttt{linalg}-based generated assembly, as the flow of instruction generation in the simulator is hard coded.

For the \emph{upmem} configuration we use a high-end real UPMEM machine with 16 DIMMs.
Each DPU runs at \SI{350}{\mega\hertz} with  \SI{64}{\mega\byte}
of main RAM (MRAM) and a \SI{64}{\kilo\byte} of working RAM (WRAM).
For the \emph{hbmpim} configuration, we use Samsung’s simulator~\cite{samsungOneMCC} with its default settings. The HBM-PIM system comprises 16 parallel channels, each containing 16 banks and 8 PIMUnits per channel. Each PIMUnit includes 16 floating-point units (FPUs), resulting in a total system capacity of \SI{6}{\giga\byte}.
Accurately modeling the data communication time over the off-chip bus of the host processor is beyond the scope of this work and has been excluded from all experiments. 

For each benchmark (see Sec.~\ref{subsec:bench}), 1024 random maps are generated and all results reported in this paper correspond to the geometric mean of ten execution runs.
\costmodelframework and UPMEM's simulators are ran on an AMD Ryzen 9 7900X3D @\SI{4.4}{\giga\hertz} CPU having a maximum
clock frequency of \SI{5.6}{\giga\hertz}.
The machine has \SI{64}{\giga\byte} of main
memory (DRAM) and runs a Linux Ubuntu (22.04).
\subsubsection{Benchmarks}
We use two commonly used set of benchmarks to evaluate the UPMEM and HBM-PIM systems.
\label{subsec:bench}
\\\noindent\textbf{UPMEM benchmarks:}
We evaluate on the open-source PrIM benchmark suite~\cite{cnm-benches}. 
\textit{Vector addition (va)} represents element-wise addition of two input vectors. \textit{Matrix-vector multiply (gemv)} is a dense linear algebra application multiplying a matrix with a vector. 
\textit{Select (sel)} is a database operator that, given an input array,
filters the array elements according to a given input predicate. 
\textit{Image histogram (hst)} calculates the histogram of an image,
i.e., it counts the number of occurrences of each possible pixel value
in an input image and stores the aggregated counts of occurrences
into a set of bin. 
\textit{Scan (scan)} processes an input array by performing a prefix operation (such as addition) to produce a cumulative result for each element and modifies the array in-place.
\textit{Matrix-matrix multiplication (gemm)}
is a ubiquitous compute kernel in numerical computing in
general and machine learning in particular. 
\textit{Reduction (red)} computes the sum of the elements in an
input array. 
\\\noindent\textbf{HBM-PIM benchmarks:}
Since HBM-PIM is ML-specific and can not run non-ML workloads, we evaluate it using the set of benchmarks from the original paper~\cite{samsungPIM} plus two additional ML workloads.
The set includes \emph{gemv}, \emph{va}, \emph{ReLU (relu)}, a non-linear activation function, \emph{Google’s RNN Transducer (rnnt)}~\cite{google_rnnt}, \emph{AlexNet(alexnet)}~\cite{alexnet},  and \emph{ResNet(resnet)}~\cite{resnet}, two widely used computer vision models, and \emph{Wav2(wav2)}~\cite{wav2}, a self-supervised speech representation model.

\subsection{\costmodelframework validation}
To validate \costmodelframework, we compared its estimated latencies against  execution latency on UPMEM hardware and the simulated latency on HBM-PIM simulator. For each benchmark, 1024 random mappings for different application with different iteration spaces were generated and executed.
\subsubsection{UPMEM latency prediction}
\label{subsec:validation}
\input{figures/evaluation/upmem_validate_small}

Fig.~\ref{fig:upmem_validation} illustrates the accuracy by comparing the absolute predicted execution times with those measured on real hardware for various kernels and configuration modes. 
As shown in the figure, data-intensive kernels, such as \textit{sel}, \textit{hst}, \textit{red}, and \textit{va} are dominated by DMA instructions latency and exhibit a higher average error of 8.9\%, primarily due to the unpredictable nature of the DMA engine. 
Conversely, compute-intensive kernels dominated by arithmetic instructions such as \emph{gemv} and \emph{gemm}, show a lower average error of 5.5\%. The \textit{hst} kernel lies between these two categories, balancing the latency ratio, with an average error of 6.3\%.
In most kernels, the assembly code generated from C shows a better accuracy than the generated from the DSL. However, the difference is minimal (geom difference of 1.7\%), demonstrating that the performance of the generated codes is very close to the clang-generated code.

\subsubsection{HBM-PIM latency prediction}
Fig.~\ref{fig:samsung_validate} shows the latency prediction accuracy of \costmodelframework across different benchmarks, compared to the HBM-PIM simulator. The average prediction errors for the simple kernels—\emph{gemv}, \emph{va}, and \emph{relu}—are 3.04\%, 5.89\%, and 0.90\%, respectively. \emph{relu}, being the simplest kernel with the fewest instructions, exhibits the lowest prediction error. In contrast, \emph{va} shows a higher error than \emph{gemv}, likely due to more frequent instruction transitions across odd and even banks, which introduces more timing complexity in hardware. \emph{gemv}, with fewer inter-bank instruction jumps, demonstrates more stable timing behavior.
For \emph{gemv} and \emph{relu}, there is a significant gap between maximum and average errors, suggesting that the average error is closer to the minimum. This is not observed in \emph{va}, indicating greater variability in its prediction accuracy. For these simple kernels, varying the kernel size allowed us to generate numerous variants. In contrast, the more complex benchmarks (\emph{rnn-t}, \emph{alexnet}, \emph{resnet}, and \emph{wav2}) have fixed layer sizes, limiting the number of kernel variants and leading to a smaller gap between minimum and maximum errors.
Overall, \costmodelframework achieves an average prediction error of 2.99\%, with a maximum error of 5.78\% and a minimum error of 0.40\%.

\input{figures/evaluation/samsung_validate}

\subsection{Comparison to existing CNM evaluation frameworks}
\label{subsec:eval-soa-comparison}
In this section, we compare the results of \costmodelframework against other available CNM frameworks. Unfortunately, for the HBM-PIM target, no other CNM framework currently exists that models the system in sufficient detail. 
Focusing on UPMEM, Table~\ref{tab:break_down} compares \costmodelframework's performance (accuracy and estimation time) for codes generated from linalg and C, (\emph{our-metric(c)} and \emph{our-metric(gen)}) with the UPMEM simulator (\emph{sim}) across three metrics: prediction time relative to the actual hardware (\emph{speedup}), worst-case error in estimation (\emph{worst}), and percentage of runs with an error margin below 15\% (\emph{g-res}).
\input{figures/evaluation/result_breakdown_2}
As shown in the table, \costmodelframework's execution time (our-speedup) is comparable to the real hardware, averaging 0.638$\times$ of the actual hardware, while being up to 8620000$\times$ faster than the UPMEM's simulator. The UPMEM simulator is, on average, seven orders of magnitude slower than the hardware.

As shown in the table, both generated codes, in their worst cases, across different kernels achieve values relatively close to the hardware.
Among the kernels in both versions, \costmodelframework has the worst performance on the \emph{red} and \emph{va} kernels, as these are very memory-intensive and can have unpredictable memory interaction. On the other hand, \costmodelframework performs significantly better, as our results show that the UPMEM simulator is often inaccurate. Across the kernels, for the same reasons, a similar trend in the worst accuracies can be seen in the UPMEM simulator, especially in data-intensive benchmarks.
\costmodelframework predicts accurately 86.5\% of the time using assembly from Clang, and 85.3\% of the time using code generated from \linalg. On the other hand, UPMEM’s simulator predicts correctly only 29.8\% of the time. This is because it does not accurately simulate DMA requests and assumes a constant execution time for all of them, regardless of type or size. The uPIM~\cite{upim} simulator has the same limitation, as it also cannot model DMA timing accurately.

\subsection{Design space exploration}
\label{subsec:eval-dse}
In this section, we evaluate ``what-if'' scenarios by relaxing the constraints imposed by the CNM systems. We specifically focus on UPMEM (as the results can be compared to real system) and evaluate how system modifications can improve application performance for specific domains. This helps in understanding the system-level changes needed to achieve desired results for certain workloads. We demonstrate that increasing resources does not necessarily benefit all applications equally. Using \costmodelframework, we model a hypothetical system (\emph{modeled}) and compare its performance against the standard \emph{upmem} configuration. All experiments assume 16 tasklets per DPU. The results, shown in Fig.~\ref{fig:dse}, highlight the performance variations for different benchmarks under the following scenarios. 
This is particularly beneficial in situation if someone wants to see how should the system be changed for applications from a particular domain to get the desired results. The idea is also to demonstrate that by changing the system, e.g., by increasing the compute units or other resources, not all applications can benefit from it. We model the hypothetical system in \costmodelframework (\emph{modeled}) and compare its performance to the \emph{upmem} configuration and present the results in Fig.~\ref{fig:dse}. Note that the UPMEM system is the same as in Sec.~\ref{subsec:setup} and the performance is expected to be different. For all experiments, we use 16 tasklets per DPU. 
~\\\noindent \textbf{DMA engine:}
Kernels such as \emph{va} might have DMA contention due to smaller arithmetic intensity. We model a system where WRAM and MRAM are connected via multiple ports, and data communication between them take a constant time. For vector sizes of $2^{32}$ (type INT32), the figure shows that the difference in the results is less than 1\%, indicating that \textbf{proper mapping}, i.e., correctly assigning WRAM to tasklets, can mitigate the impact of DMA.
~\\\noindent \textbf{Execution pipeline:}
The minimum cycles per instruction (CPI) for a tasklet is 11 cycles due to the hardware constraints (see Sec~\ref{upmem-background}). If we reduced the pipeline to 4 stages, then utilizing four tasklets could fully utilize the pipeline, resulting in a CPI of 4 cycles. As depicted in the figure, an \emph{mv} kernel of size  $2^{15}\times2^{14}$ running on such a system achieves a $1.7\times$ speedup compared to \emph{upmem}. This is because each tasklet gets a larger WRAM chunk (WRAM size is unchanged) which reduces the number of DMA requests.
\input{figures/evaluation/dse}

~\noindent \textbf{Functional units:}
DPUs lack dedicated functional units, e.g., floating-point units and multipliers, and these operations are emulated in software. Consequently, their latencies are value-dependent, which might be advantageous in certain scenarios (e.g., applications using smaller values). Our evaluations show an average CPI of 112 for data values ranging between (0 and 32), and the number of cycles increases as the values increase. We modeled DPUs having dedicated multipliers that take 44 cycles. On the \emph{mm} kernel with $2^{10} \times 2^{10}$ size matrices, the \emph{modeled} system outperforms \emph{upmem} by $2.18\times$.
Similarly, by doubling the WRAM size, i.e., \SI{128}{\kilo\byte}, kernels like \emph{red} $1.97\times$ speedup on the \emph{modeled} system compared to \emph{upmem}. Contrarily, doubling the MRAM size to \SI{128}{\mega\byte} does not any show any performance difference for the  \emph{hst} benchmark.
~\\\noindent \textbf{WRAM size:}
Because of the WRAM's limited size, increasing the number of tasklets results in smaller WRAM allocations for each tasklet. We modeled DPUs with 128KB of WRAM and compared the execution of the \emph{red} kernel with an input size of $2^{32}$. \emph{modeled} shows approximately $1.97\times$ speedup (see Fig.~\ref{fig:dse}).
~\\\noindent \textbf{MRAM size:}
Each DPU has direct access to only 64MB of MRAM, imposing a constraint on the amount of data available for each execution. To assess this limitation, 
We model an UPMEM with \SI{128}{\mega\byte} MRAM and compare results for the \emph{hst} benchmark with a data size of $2^{31}$ on real hardware and on a modeled system where each DPU has access to 128MB of MRAM. In this kernel, the results indicate no change in the execution time of the kernel on the system. %

\begin{figure}
\centering
\includegraphics[width=\columnwidth]{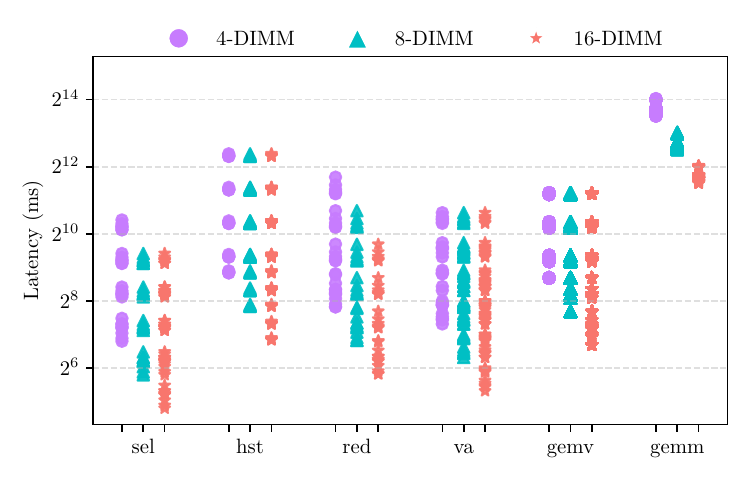}
\caption{Exploring different mappings of different kernels using \costmodelframework.}
\label{fig:distribution_plot}
\vspace{-5pt}
\end{figure}

\subsection{Mapping space exploration}
The rapid estimation time of \costmodelframework enables exploration of the huge design and transformation spaces. 
Fig.~\ref{fig:distribution_plot} shows the predicted latency for various kernels across different configurations: 4-DIMMs (8 ranks), 8-DIMMs (16 ranks), and 16-DIMMs (32 ranks). For each kernel, the same iteration space as in \cite{cinm} was used. All possible mappings were exhaustively generated and predicted.

As shown in the figure, manual decisions based on educated intuitions such as increasing the number of DIMMs for improved performance can easily lead to performance degradation. %
For example, in the \emph{va} kernel, 65.47\% of the mappings on 16-DIMMs perform worse than the best mapping on 4-DIMMs, and 85.71\% perform worse than the best mapping on 8-DIMMs. This trend is observed across the majority of evaluated applications, underscoring the necessity for precise modeling tools like \costmodelframework that accurately capture and consider the interplay of all CNM system parameters. %

%% file: figures/evaluation/upmem_validate_small.tex
\begin{figure}[b]
    \footnotesize
    \pgfplotsset{compat = newest}
    \pgfplotsset{major grid style={dotted,aluminium2!50!black}}
    \begin{tikzpicture}
    \begin{axis}
    [
        width=\columnwidth,
        ybar=1pt, %
        enlargelimits=0.08,
        enlarge y limits={upper, value=0.1},
        ylabel style={align=center},
        y label style={at={(-0.065,0.5)}},
        ylabel= Error (\%),
        legend style={draw=none, fill=none},
        bar width=5pt,
        legend columns=7,
        height=0.22\textwidth,
        ymajorgrids=true,
        grid style=dashed,
        axis x line*=bottom,
        x tick label style={xshift=.0em, yshift=-.0em, rotate=45,anchor=east},
        yminorticks=true,
        legend style={at={(0.98,0.92)},anchor=north east},
        xlabel={},
        ytick scale label code/.code={\pgfmathparse{int(-#1)}$y \cdot 10^{\pgfmathresult}$},
        every y tick scale label/.style={at={(yticklabel cs:0.5)}, anchor = south, rotate = 90},
        symbolic x coords={sel, hst, scan, red, va, gemv, gemm, geom},
        xtick=data,
    ]

    \addplot+ [blind_safe_five_scheme_seven_colors_grnblu] coordinates {
    (sel, 9.502332) 
    (hst, 4.85995) 
    (scan, 9.845124)
    (red, 10.27539) 
    (va, 9.148598) 
    (gemv, 4.626080) 
    (gemm, 5.4636825)
    (geom,7.28)
    }; 

    \addplot+ [blind_safe_six_scheme_seven_colors_grnblu] coordinates {
    (sel, 8.05558) 
    (hst, 8.13124) 
    (scan, 10.14123)
    (red, 11.20965) 
    (va, 11.35297) 
    (gemv, 6.228443) 
    (gemm, 6.4475634)
    (geom, 8.795)
    };

    \legend{c, gen}
    \end{axis}
    \end{tikzpicture}
    \caption{\costmodelframework performance estimation error for UPMEM (compared to real hardware (upmem)). The figure also compares the code generation for different input application representations (C vs linalg).}
    \label{fig:upmem_validation}
\end{figure}
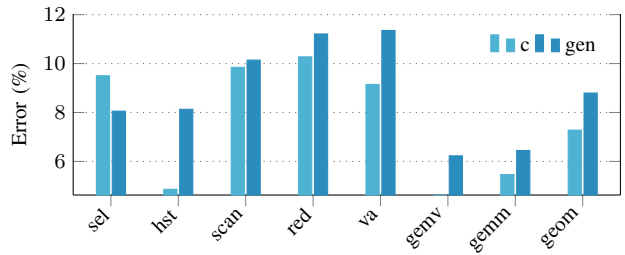

%% file: figures/evaluation/samsung_validate.tex
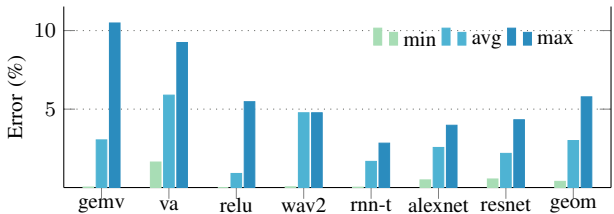
\begin{figure}[t!]
    \footnotesize
    \pgfplotsset{compat = newest}
    \pgfplotsset{major grid style={dotted,aluminium2!50!black}}
    \begin{tikzpicture}
    \begin{axis}
    [
        width=\columnwidth,
        ybar=1pt, %
        enlargelimits=0.08,
        enlarge y limits={upper, value=0.1},
        ylabel style={align=center},
        y label style={at={(-0.055,0.5)}},
        ylabel= Error (\%),
        legend style={draw=none, fill=none},
        bar width=4pt,
        legend columns=7,
        height=0.22\textwidth,
        ymajorgrids=true,
        grid style=dashed,
        axis x line*=bottom,
        x tick label style={xshift=.0em, yshift=-.3em, rotate=0,anchor=center},
        yminorticks=true,
        legend style={at={(0.55,0.92)},anchor=north west},
        xlabel={},
        ytick scale label code/.code={\pgfmathparse{int(-#1)}$y \cdot 10^{\pgfmathresult}$},
        every y tick scale label/.style={at={(yticklabel cs:0.5)}, anchor = south, rotate = 0},
        symbolic x coords={gemv, va, relu, wav2, rnn-t, alexnet, resnet, geom},
        xtick=data,
    ]

    \addplot+ [blind_safe_three_scheme_seven_colors_grnblu] coordinates {
    (gemv, 0.0407966)
    (va, 1.632653061)
    (relu, 0.010821046936291087)
    (wav2, 0.054615)
    (rnn-t, 0.03294903157)
    (alexnet, 0.4966741408)
    (resnet, 0.5549852923)
    (geom, 0.403356317)
    }; 

    \addplot+ [blind_safe_five_scheme_seven_colors_grnblu] coordinates {
    (gemv, 3.0427736)
    (va, 5.890094094505)
    (relu, 0.9057534288069706)
    (wav2, 4.76897160)
    (rnn-t, 1.67413761557)
    (alexnet, 2.560668928)
    (resnet, 2.18381561590)
    (geom, 2.999831673)
    }; 
    
    \addplot+ [blind_safe_six_scheme_seven_colors_grnblu] coordinates {
    (gemv, 10.481538199761783)
    (va, 9.239044454)
    (relu, 5.472086035109916)
    (wav2, 4.76897160)
    (rnn-t, 2.83514461101)
    (alexnet, 3.9736072775972)
    (resnet, 4.32421720174)
    (geom, 5.782191357)
    };

    \legend{min, avg, max}
    \end{axis}
    \end{tikzpicture}
    \caption{Accuracy of predicted latency for Samsung's HBM-PIM (minimum, average, and maximum absolute error). }
    \label{fig:samsung_validate}
\vspace{-7pt}
\end{figure}

%% file: figures/evaluation/result_breakdown_2.tex
\begin{table}[tb]
\footnotesize
\caption{\costmodelframework accuracy and speedup comparison with existing simulation frameworks}
\begin{center}
\begin{tabular}{>{\ttfamily}p{0.26\columnwidth}p{0.05\columnwidth}p{0.05\columnwidth}p{0.05\columnwidth}p{0.05\columnwidth}
p{0.05\columnwidth}p{0.05\columnwidth}p{0.05\columnwidth}}
\toprule
\sffamily \textbf{Metric} & sel & hst & scan & red & va & gemv & gemm \\ 
\midrule
\sffamily our-speedup(c) & .696 & .742 & .721 & .780 & .479 & .065 & .593  \\ 
\sffamily our-speedup(gen) & .75 & .793 & .788 & .862 & .436 & .083 & .792  \\ 

\sffamily sim-speedup & 8.9$\,\mu$ & 300$\,\mathrm{n}$ & $100\,\mathrm{n}$ & $100\,\mathrm{n}$ & 300$\,\mathrm{n}$ & 300$\,\mathrm{n}$ & 300$\,\mathrm{n}$ \\ 
\sffamily our-worst(c)(\%) & 32.6 & 20.5 & 33.1 & 30.6 & 36.3 & 21.8 & 22.9  \\ 
\sffamily our-worst(gen)(\%) & 28.7 & 27.0 & 39.3 & 44.0 & 40.3 & 27.3 & 26.5  \\ 
\sffamily sim-worst(\%) & 93.6 & 92.5 & 110 & 131 & 93.1 & 92.9 & 91.5\\ 
\sffamily our-g-res(c)(\%) & 77.1 & 97.6 & 81.2 & 76.6 & 82.6 & 96.6 & 95.9 \\ 
\sffamily our-g-res(gen)(\%) & 86.1 & 83.8 & 78.1 & 75.7 & 89.5 & 92.4 & 91.3 \\ 

\sffamily sim-g-res(\%) & 25.2 & 42.3 & 18.2 & 12.5 & 25.3 & 41.6 & 43.6 \\ 
\bottomrule
\end{tabular}
\end{center}
\label{tab:break_down}
\vspace{-7pt}
\end{table}

%% file: figures/evaluation/dse.tex
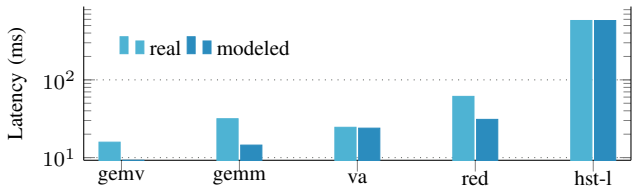
\begin{figure}[t]
    \footnotesize
    \pgfplotsset{compat = newest}
    \pgfplotsset{major grid style={dotted,aluminium2!50!black}}
    \begin{tikzpicture}
    \begin{axis}
    [
        width=\columnwidth,
        ybar=1pt, %
        enlargelimits=0.08,
        enlarge y limits={upper, value=0.1},
        ylabel style={align=center},
        y label style={at={(-0.09,0.5)}},
        ylabel= Latency (ms),
        legend style={draw=none, fill=none},
        bar width=8pt,
        legend columns=7,
        ymode = log,
        log basis y={10},
        height=0.2\textwidth,
        ymajorgrids=true,
        grid style=dashed,
        axis x line*=bottom,
        x tick label style={xshift=.0em, yshift=-.3em, rotate=0,anchor=center},
        yminorticks=true,
        legend style={at={(0.05,0.85)},anchor=north west},
        xlabel={},
        ytick scale label code/.code={\pgfmathparse{int(-#1)}$y \cdot 10^{\pgfmathresult}$},
        every y tick scale label/.style={at={(yticklabel cs:0.5)}, anchor = south, rotate = 0},
        symbolic x coords={gemv, gemm, va, red, hst-l},
        xtick=data,
    ]

    \addplot+ [blind_safe_five_scheme_seven_colors_grnblu] coordinates {
    (va, 24.529)
    (gemv, 15.79117771)
    (gemm, 31.605667)
    (hst-l, 579.884)
    (red, 61.282333)
    }; 

    \addplot+ [blind_safe_six_scheme_seven_colors_grnblu] coordinates {
    (va, 23.907) 
    (gemv, 9.235934286)
    (gemm, 14.474)
    (hst-l, 579.272333)
    (red, 31.006667)
    };

    \legend{real, modeled}
    \end{axis}
    \end{tikzpicture}
    \caption{Design space exploration using \costmodelframework.  }
    \label{fig:dse}
\vspace{-9pt}
\end{figure}

%% file: contents/conclusion.tex
\section{Conclusions}
\label{sec:conclusions}
We introduce \costmodelframework, a cost modeling framework for CNM systems that accurately estimates execution time within milliseconds, given an input application, target description, and mapping specification. 
To ensure compatibility with existing CNM compilation and DSE frameworks, we introduce llvcnm and CNM IR, two intermediate representations at different abstraction levels, that are compatible with \costmodelframework, and have also implemented a front-end that translates input applications from clang-assembly and MLIR-linalg into our intermediate representations. Furthermore, we introduce a flexible and hierarchical mapping specification that captures the structure of CNM systems. Our evaluation of \costmodelframework across a range of applications and configurations validates the accuracy of its predictions when compared against CNM simulators and a real CNM system. In contrast to existing simulators, which often struggle to accurately model the memory system and interaction with the compute units, \costmodelframework provides consistent and reliable execution time estimates, particularly for memory-intensive workloads, are is seven orders of magnitude faster compared to existing simulators.